\documentclass[prb,twocolumn]{revtex4-1} 

\usepackage{amsmath}  

\usepackage{amsfonts} 

\usepackage{graphicx} 

\usepackage{multirow} 

\usepackage{array}

\usepackage{setspace}

\begin{document}


\title{Learning Concepts First -- A Course Structure with Improved Educational Outcomes in the Short, Medium, and Long Terms (Especially for Minority Groups Underrepresented in Physics)}

\author{D. J. Webb}
\email{webb@physics.ucdavis.edu} 
\affiliation{Dept. of Physics, University of California, Davis, CA, 95616}


\date{\today}

\begin{abstract}
An active learning physics course (treatment) was re-organized in an attempt to increase students' problem solving abilities.  This re-organized course covered all of the relevant concepts in the first 6 weeks with the final 4 weeks spent in practice at solving complicated problems (those requiring students to use higher order cognitive abilities).  A second active learning course (control) was taught in the same quarter by the same instructor using the same curricular materials but covering material in the standard (chapter-by-chapter) order.  After accounting for incoming student characteristics, students from the treatment course scored significantly better than the control for two outcome measures: i) the final exam and ii) their immediately subsequent physics course.  More importantly, students from minority groups who are underrepresented in physics had final exam scores as well as class grades that were indistinguishable from the rest of their class if and only if they were in the treatment class.  Finally, many of the students in this cohort took a Concepts First course in their third quarter of introductory physics.  The students who took at least one Concepts First course are found to have significantly higher rates of graduation with a STEM major than those students from this cohort who did not take a Concepts First course. 
\end{abstract}

\maketitle 

\section{Introduction} 

Almost 30 years of experience in teaching physics along with a few years of reading about Physics Education research \cite{PERResearch} had left this author without a satisfactory personal understanding of how the issues of learning physics concepts versus learning to solve physics problems are connected and of how to teach them.  Research\cite{SimonChess, ChiExpert} suggests that an expert problem solver has developed a large and well-structured knowledge in the field of their expertise so a teacher's job is presumably to help their students begin to build this knowledge structure.  A ``large'' knowledge must come from practice analyzing and solving a large and varied set of problems, but the fact that this large knowledge is ``well-structured'' implies either that solving many problems should happen concurrently with learning concepts or that the concepts should be largely in place early-on so that the practice solving many problems can meaningfully deepen and strengthen an existing conceptual knowledge structure.  Most research seems to conclude that concepts are not easily learned \cite{SwellerCogLoad} by simply working problems and may not be learned at all\cite{1000Problems} even after working many problems.  There is even research\cite{PerryElemSch} that suggests, to this author, that learning to work problems can actually impede the development of the appropriate conceptual understanding.  If it is true that an appropriate conceptual knowledge structure is not developed in working problems then one might try the second possibility, build a basic conceptual knowledge framework first and then strengthen this framework with practice analyzing and solving complicated
problems.\cite{VanHeuv}   An experiment along these lines was carried out a few years ago in introductory physics here at UC Davis and is discussed in this paper.

\section{Method}
In a recent Spring Quarter UC Davis offered four lecture sections, Sections 1 through 4, of Physics 9A (classical mechanics).  A student in Physics 9A has three hours of lecture, one hour of discussion section, and 2.5 hours of laboratory each week. \ The textbook and the laboratory experiments for all four sections were exactly the same but the lectures, discussion sections, and homework are under the control of the instructor assigned to teach that section.  The textbook used in all lecture sections was ``University Physics'' ($13^{th}$ ed.) by Young and Freedman and each lecture section covered Chapters 1-11 and 13-14 (including translational motion, force and Newton's Laws, energy, momentum, torque and rigid
body motion, angular momentum, Newton's Law of Gravity, and oscillations).\cite{Coverage}  The material in Section 1 was organized so that students worked to learn the ideas over the first 6 weeks of the 10-week quarter and then used the final 4 weeks to apply those principles to solve the analytically (and algebraically) complicated physical problems that physicists prize.  This course used active learning techniques so this treatment group will be called ``Active Learning Plus Concepts First'' (AL\&CF).  The other three sections learned ideas at the same time as calculations over the entire 10 weeks of the quarter.  Section 2 was also taught using active learning techniques and so will be called ``Active Learning Only'' (ALOnly).  Sections 1 (AL\&CF) and 2 (ALOnly) were both taught by the author and had identical homework problems,\cite{Mastering} discussion section problems/questions, lab problems, and lecture questions.\cite{MaterialsOnline}  These two sections were active learning classes in that there were peer-peer discussions happening in both the lecture (in answering clicker questions) and in the one-hour per week discussion section (where attendance was required) and there were conceptual questions on midterm exams.

In practice, lecture-time for these two active-learning sections ran in a fairly standard way.  A new topic was motivated either by a demonstration or by a discussion of a real world situation and continued with a presentation of relevant physical ideas expressed in English and in equations but also using diagrams and graphs as appropriate.\cite{Learning}  After about 10-20 minutes of lecture there would be time for 2-5 clicker questions that test the students understanding of the basic concepts and/or help the students to build this basic understanding.  Usually students talk to each other about the clicker question (Peer-Peer discussion) for a couple of minutes before they are polled and sometimes the students are polled before they can discuss the issue between themselves and then polled again after they have discussed it, a la Mazur's prescription from Peer Instruction.\cite{Mazur}  After student polling, the lecturer usually either explains the physics involved in answering the question or conducts a whole-class discussion.  Many of the questions used were directly from Mazur's book but some were either written by the author or borrowed from other UC Davis instructors. \ After this kind of conceptual discussion and
clicker questions, a lecture from section 1 would likely move on to another concept but a lecture from section 2 would more likely move on to using the ideas in equation form to analyze/explain a physical situation (i.e. work a practice problem).  During the 1-hour per week discussion section students worked in groups of 4-5 to answer qualitative questions often concerned with their homework problems.  Examples of these can be found in the course materials.\cite{MaterialsOnline}  A graduate TA moved from group to group, offering help as necessary and conducting a whole-class discussion after most groups finish a question.  Finally, midterm exams included a mixture of qualitative and quantitative problems as appropriate for the particular exam (again, see Ref. \onlinecite{MaterialsOnline}).

At this point I should note that I will be using a somewhat strict definition of the term ``Active Learning'' by including only classes where the lecturer involves the students in peer-peer discussions on conceptual issues and where the exams include one or more questions that are primarily conceptual.  Sections 1 and 2 are the focus of this paper but some of the data will necessarily involve students from the other Sections 3 and 4 so we should note that the instructors in the other two sections historically received significantly higher student evaluations than the instructor in Sections 1 and 2 but that Sections 3 and 4 were not active learning classes (i.e. they were NoAL courses) in that neither used active learning techniques in lecture or discussion although Section 3 did include some conceptual questions on exams.

In Winter Quarter following the original experiment this cohort of students took Physics 9C, a course which covers electricity and magnetism (including electric circuits).  The author taught one of the four sections of 9C and ordered the course in the same way as the treatment course of 9A (for electricity and magnetism the complicated two and three dimensional integrations were among the problems dealt with in the last 4 weeks).  In addition, another of the four instructors (one who always runs an active learning course) was recruited to organize his course in the same way; 6 weeks of conceptual learning followed by 4 weeks of calculations.  Thus, the original cohort (made up of all 4 lecture sections) of 9A students had many who had experienced an AL\&CF class and some who had taken two AL\&CF classes, some who had ALOnly but no AL\&CF (there were ALOnly courses in 9B during the intervening quarter also), and the rest had only NoAL classes.  There was no controlled study during this quarter but since the original experiment was done almost 4 years ago there is now graduation data available for this cohort of students.

The students had no knowledge of how any of the sections would be taught when they enrolled.  In addition, each class was completely full and almost no students could switch sections after they heard how the class would be taught so, although this is not a randomized trial, student selection issues related to the course structure are probably negligible even though student selection regarding instructor is likely.  Table \ref{IncomingData} shows a number of measurements of student academic skills and attitudes as they enter the course (the errors shown are standard deviations showing the breadth of the distributions).  The American Physical Society\cite{APSURM} notes that women and a set of minority groups (the variable UndRpMns\cite{Hispanic} includes African Americans, Native Americans, and Hispanic Americans) are underrepresented in physics so this paper will discuss results for women and the group of students (self-identifying as) belonging to one of these underrepresented minority groups along with the entire class.  FCI refers to the Force Concept Inventory,\cite{FCI} which is a 30-question multiple-choice survey of the student's understanding of forces, motion, and their Newtonian relationship.  This survey was given to all students at the beginning of the quarter (pretest) and again (posttest) in the 8th week of the 10-week quarter at least 3 weeks after every section had finished working on the Newtonian physics of point-like objects.  We take the pretest (FCIpre) to be a measure of the initial Newtonian understandings that the students bring to the class.  CLASS refers to the Colorado Learning Attitudes about Science Survey\cite{CLASS} which was given both at the beginning of the course (pre) and at the end of the course (post) but was only given to the two sections that are the main focus of this paper.  The CLASS questions have 5-point Likert-scale.\cite{Likert}  Each question is scored as either the same as an expert would answer (favorable), neutral, or opposite to how an expert would answer (unfavorable).  To produce a single score\cite{OtherCLASS} we use the fraction of favorable answers minus the fraction of unfavorable answers so a higher score is associated with a more expert-like attitude/epistemology.  A student's GPA is their UC Davis GPA at the beginning of the quarter and we take this to be a measure of their general academic ability.  To control for math skill we wanted to use an average of a student's introductory calculus grades but examination of the grade distributions in these math classes suggested that different instructors end up with very different grade distributions in their classes (suggesting very different grading decisions).  We decided to account for these instructor decisions by normalizing the grade distribution for each course each quarter it was given so that each course had a grade distribution with average grade = 0 and a standard deviation of 1 each quarter (in other words, we converted each course's grades to z-scores quarter by quarter).  Then IntrMathZ is the average of a student's z-scores in their introductory calculus courses completed by the end of the spring quarter\cite{MathScores} in which they took the physics courses described in this paper.  You will notice that as a group the students in these physics courses had math scores about 0.2 to 0.4 standard deviations above the average in their respective classes.  ``Semesters Physics'' refers to the number of semesters of previous physics classroom experience and was self-reported by the students.

\begin{table*}
\caption{\label{IncomingData}Five incoming measures for these three course-types. \ Sections 3 and 4 (NoAL with about 330 students in the Class, about 90 Females, and about 50 UndRpMn, each of these depending slightly on the
specific measurement) are grouped together. \ Section 1 (AL\&CF) has about 155 in the Class, about 38 Females, and about 18 UndRpMns and Section 2 (ALOnly) has about 160 in the Class, about 40 Females, and about 20 UndRpMns. \ Shown
are distribution averages (standard deviations) for the five incoming measures of student abilities or learning attitudes.}
\begin{tabular}{| >\centering m{50pt} | >\centering m{65pt} | >\centering m{60pt} | >\centering m{70pt} | >\centering m{60pt} | >\centering m{60pt} | m{60pt} <\centering |}
\hline Lecture  &  Group &  GPA  &  IntrMathZ  & FCI$_{pre}$ & Semesters Physics & CLASS$_{pre}$ \\ \hline \hline
1  &  Class.  & 2.96 (0.57)  & 0.17 (0.69)  & 16.1 (7.0) & 1.9 (1.0) & 0.47 (0.22) \\ \cline{2-7}
AL\&CF &  UndRpMns  & 2.69 (0.60)  & -0.12 (0.73)  & 15.6 (5.7) & 2.0 (0.8) & 0.52 (0.17) \\ \cline{2-7}
 &  Females  & 3.03 (0.56)  & 0.21 (0.67)  & 12.0 (6.5) & 1.6 (1.0) & 0.42 (0.24) \\ \hline
2  &  Class  & 2.96 (0.52)  & 0.11 (0.67)  & 16.3 (6.4) & 1.9 (1.2) & 0.46 (0.23) \\ \cline{2-7}
ALOnly &  UndRpMns  & 2.70 (0.37)  & -0.11 (0.49)  & 14.2 (5.6) & 1.7 (1.0) & 0.47 (0.22) \\ \cline{2-7}
 &  Females  & 3.01 (0.55)  & 0.19 (0.76)  & 13.1 (6.0) &  1.5 (1.1) & 0.42 (0.21) \\ \hline
3  &  Class  & 3.11 (0.58)  & 0.40 (0.67)  & 15.5 (6.8) & 1.8 (1.2) &  \\ \cline{2-7}
NoAL  &  UndRpMns  & 2.70 (0.63)  & 0.03 (0.68)  & 13.7 (6.4) &  1.6 (1.0) & \\ \cline{2-7}
 &  Females  & 3.12 (0.51)  & 0.38 (0.62)  & 12.1 (6.0) & 1.7 (1.2) &  \\ \hline
\end{tabular}
\end{table*}

\section{Results}

We have three main immediate outcome measures: the final exam, the FCI given in $8^{th}$ week of the class to all four lecture sections, and the CLASS given the $9^{th}$ week of the class only to Sections 1 and 2.  First we'll briefly point out that the CLASS survey showed no significant differences between Sections 1 and 2.  The two sections had similar CLASSpre scores that were not appreciably changed by the course.  We will deal with FCI gains after discussing the final exam scores.  A little more detail on each part of the ensuing statistical analysis is included with the course materials.\cite{MaterialsOnline}

\subsection{Final Exam}

Each of the four lecture sections took the same final exam at the same time (the final exam is included in the course materials\cite{MaterialsOnline}).  This exam was written by the two instructors from sections 3 and 4 with additional help from a third instructor who often teaches this course.  The instructor in Sections 1 and 2 had no input on the final and did not see it until all instruction (including review sessions) had ended.  There were eight problems on the final exam and they had equal value.  For Sections 1 and 2 together, each final exam problem was graded a single grader.  The instructor from Section 3 supervised the grading.

We normalized the final exam distribution by calculating z-scores using the average and standard deviation for the entire group of students from all four lecture sections.  To assess the value of the particular curriculum organization of Section 1 (treatment group is AL\&CF class), we can control for the student-level effects of 1) general academic skill using incoming GPA, 2) mathematical ability using introductory math scores, 3) previous understanding of Newtonian mechanics using FCIpre, 4) previous physics experience using Semesters Physics, and 5) attitudes toward learning physics using CLASSpre.  Comparing AL\&CF with ALOnly for each of the three groups (Class, UndRpMns, and Females) identified we find that each group had higher final exam scores when they were in the AL\&CF course.  This is true whether or not we control for students' GPA's, math skills, etc.  However, controlling for those variables reduces the error estimates for the comparison between AL\&CF and ALOnly so we control for these student-level variables. \ \textbf{We find that the AL\&CF Class had higher final exam z-grades by $0.18 \pm 0.07$} (from now on the errors quoted are standard errors), \textbf{the AL\&CF UndRpMns had higher final exam z-grades by $0.92 \pm 0.23$}, and the AL\&CF Females had higher final exam z-grades by $0.13 \pm 0.14$.  The first two of these numbers would generally be considered statistically significant in that error estimate suggests that there is only about a 2\% chance of getting these data for the whole Class if there was actually no effect at all and less than 0.01\% chance of getting these data for UndRpMns if there was no effect at all.  A survey\cite{EffectSizes} of effect sizes in education experiments shows us that an average size effect for a whole-class education experiment is about $0.18 \pm 0.41$ standard deviations. \ For this reason, we would describe the effect of AL\&CF for the whole Class as a medium size positive effect and the effect of AL\&CF for the underrepresented minorities group as a large positive effect.  Finally, we note that every identifiable group with enough statistical power (this includes Males, Chinese Americans, and East Indian Americans as well as Females) showed effects with the same sign as those for the whole class and some of those results reached statistical significance.

\subsection{Conceptual Gains}

Regarding conceptual learning, we analyze results of the (30 question) FCI by computing a normalized gain for each student as shown in equation \ref{Gain}.

  \begin{equation}
  \label{Gain}
    \substack{\text{Student} \\ \text{level gain}} =
    \begin{cases}
      \frac{FCI_{post} - FCI_{pre}}{30 - FCI_{pre}}, & \text{\small if}\ \scriptstyle FCI_{post} > FCI_{pre} \\
      \text{drop}, & \text{\small if}\ \scriptstyle FCI_{post} = FCI_{pre} = 30 \text{ or } 0 \\
      0, & \text{\small if}\ \scriptstyle 0<FCI_{post} = FCI_{pre} < 30 \\
      \frac{FCI_{post} - FCI_{pre}}{FCI_{pre}}, & \text{\small if}\ \scriptstyle FCI_{post} < FCI_{pre}
    \end{cases}
  \end{equation}
  
We compute an average gain for each section by averaging these student level gains.  The results are shown in Table \ref{FCIScores} for each of the groups we have been discussing.\cite{WCGains}  As is almost always found,\cite{Hake} classes where active learning techniques are used (Sections 1 and 2) have higher gains on conceptual surveys than the standard lecture type of class.  The FCI gain results for Class and UndRpMns in AL\&CF are statistically significantly higher than the respective gains for NoAL in that the data are less than 0.01\% likely to occur for the whole Class and roughly 0.03\% likely for UndRpMns if there were no differences between AL\&CF and NoAL.

\begin{table*}
\caption{\label{FCIScores}Average gain on Force Concept Inventory (FCI) for each group in each course-type along with the relevant standard errors.}
\begin{tabular}{| >\centering m{50pt} | >\centering m{65pt} | m{70pt} <\centering |}
\hline Lecture  &  Group &  Avg. FCI Gain \\ \hline \hline
1  &  Class.  & 0.42 $\pm$ 0.02  \\ \cline{2-3}
AL\&CF &  UndRpMns  & 0.47 $\pm$ 0.07  \\ \cline{2-3}
 &  Females  & 0.31 $\pm$ 0.05   \\ \hline
2  &  Class  & 0.37 $\pm$ 0.03   \\ \cline{2-3}
ALOnly &  UndRpMns  & 0.32 $\pm$ 0.06   \\ \cline{2-3}
 &  Females  & 0.32 $\pm$ 0.05  \\ \hline
3  &  Class  & 0.26 $\pm$ 0.02   \\ \cline{2-3}
NoAL  &  UndRpMns  & 0.19 $\pm$ 0.04   \\ \cline{2-3}
 &  Females  & 0.22 $\pm$ 0.03   \\ \hline
\end{tabular}
\end{table*}

\subsection{Transfer}

For the next comparison we describe how this cohort of students performed in their next introductory physics class, Physics 9B, which most of them took in the Fall quarter following the Spring quarter in which they took 9A.  The author had nothing to do with any of the teaching or grading in any of these 9B courses.  Physics 9B includes a diverse set of topics including waves, optics, and thermodynamics so there is very little overlap between concepts in 9A and those in 9B.  For this reason, an effect on 9B grades will be considered a transferred effect.  We compare 9B z-grades (calculated separately for each particular lecture section of 9B) controlling for the same set of student academic variables we used for the final exam fits above. Again, all of the individual groups with large enough statistical power show \textbf{AL\&CF students performing better than ALOnly students} with the whole Class scoring better by $0.22 \pm 0.09$ which is a medium size effect (these data are estimated to have about a 2\% chance of occurring if, in fact, there were no effect).  The increased 9B z-grades for Females ($0.1 \pm 0.2$) and for UndRpMns ($0.14 \pm 0.23$) in AL\&CF compared to ALOnly were not statistically significant.

\subsection{Course Grades for Minority Groups Who are Underrepresented in Physics}

It is particularly interesting that one finds no significant grade gap\cite{FExamEquit} between students from minority groups who are underrepresented in physics and the rest of the class in the AL\&CF course although there are significant (between 0.5 and 1 standard deviation) grade gaps for the ALOnly and NoAL courses. \ Because the grade distributions for the 4 courses may differ, we use z-grades for all 4 sections and measure the grade gap (ZGradeGap = CourseZGrade$_{UndRpMns}$ -- CourseZGrade$_{RestOfClass}$) between underrepresented minority groups and the rest of the class (a negative grade gap means that minority groups underrepresented in physics had a lower average grade).  We find ZGradeGap = $-0.08 \pm 0.24$ (i.e. \textbf{no significant grade gap}) for AL\&CF but ZGradeGap = $-0.88 \pm 0.22$ for ALOnly and ZGradeGap = $-0.60 \pm 0.14$ for NoAL (i.e. relatively large grade gaps).  One can get a clearer picture of the statistical importance of this finding by controlling for the incoming variables of the students in each section and estimating the chances that underrepresented minority students in the AL\&CF just randomly received higher grades.  Comparing the AL\&CF section with all three other sections and suggests that there is approximately 3\% chance that the large grade gaps in the other three lecture sections would narrow as much as in the AL\&CF section by random chance alone.

\subsection{STEM Major Retention}

As discussed earlier, many of the students in the cohort that began introductory physics during the quarter of the experiments took an AL\&CF course during the third quarter of their introductory physics series.  Since the original experiment was done almost 4 years ago there is now graduation data available on this cohort of students (normal graduation would have been 2.25 years after taking 9C and, to date, the data include students who graduated within 2.75 years of taking 9C).  We can compare the STEM graduation chances for a student who has taken at least one ``Concepts First'' introductory physics class to those of a student without any ``Concepts First'' class.  After controlling for the usual incoming student variables as before we find that members of this cohort who took at least one AL\&CF course are $1.7 \pm 0.3$ times more likely to have graduated with a STEM major in those 2.75 years after their 9C course and underrepresented minority group members were $2.7 \pm 1.3$ times more likely to have graduated with a STEM major.  The chances that these results happened even if there was no effect are 0.4\% for the whole cohort and 4\% for UndRpMns.  Finally, there were many students who took ALOnly courses but didn't take an AL\&CF course and the STEM graduation odds for these students were the same as those who took only NoAL courses.  So, the increases in STEM graduation rates seem to be associated with the concepts-first organization of a course rather than the active-learning activities.

\section{Discussion}

Changing classroom education is probably most commonly thought of as either revising curricular materials (textual and/or online pieces), modifying the structure of the learning environment (introduce peer-peer active learning), or even changing teachers. \ The experiment described in this paper suggests a type of change quite different from those just listed, indeed every effort was made to keep the treatment and control course identical in all of the ways just named. \ Nevertheless, even though one might have thought that the changes were small we find that the treatment had several significantly better outcomes and, as far as we know, no worse outcomes.  \textbf{As a class the AL\&CF students performed moderately better on the final exam (neither written nor graded by the author/instructor) as well as in their next physics course and were more likely to graduate with a STEM major.  In addition, the results for minority groups underrepresented in physics were so strikingly positive that these students were, on average, at parity with the rest of the students in this course.}  The moderate-size-effect increase on the final exam and the FCI scores were known immediately after the course.  The other results (increased scores in later classes and large increases in grades and STEM graduation rates by minority groups) were not noticed until recently.  The result is that after this experiment the author has taught 5 courses in this physics series but never thought it important to repeat the precise structure of the original course.  These later courses were taught in what could be called a ``watered-down'' version\cite{LaterClasses} that was kind of an average of an AL\&CF and an ALOnly course in that the first 8 weeks was spent on all of the conceptual issues in the course and then the last 2 weeks on the most complicated problems.

Some obvious future questions are; i) will we always find these results with UC Davis students or was this just a very unusual group of students (i.e. is there another reason why they significantly outperformed predictions based on their GPAs, math scores, and incoming physics understanding?), ii) is the method generalizable beyond the first quarter (mechanics) and, if so, what is the best way to do this, and iii) are the methods generalizable beyond UC Davis-type schools and, if so, which schools and what is the best way to do this?

\begin{acknowledgments}

The author thanks Wendell Potter, Emily West, Cassandra Paul, Mary Chessey, and the rest of the UC Davis PER group for useful discussion, comments, and suggestions.

\end{acknowledgments}


\begin{thebibliography}{99}
\bibitem{Coverage} There were two main topics (elasticity and damped \& forced oscillations) from these 13 chapters that were left out of the course as well as a few of the final chapter sections (including side-topics such as ``fundamental forces of nature'', ``rocket propulsion'', and ``black holes'').
\bibitem{Mastering} The online homework service, MasteringPhysics, was used for about 80\% of homework and the other 20\% of homework had solutions/graphs/diagrams/descriptions written down and turned in for grading.  MasteringPhysics includes the conceptual questions, simple one-step problems, and tutorial problems that were used in the first 6 weeks of the course as well as more complicated problems used in the final 4 weeks of the course.
\bibitem{MaterialsOnline} All curricular materials as well as the scheduling of these for the two different classes are available online at [URL will be inserted by AIP].
\bibitem{Learning} One could say that learning physics well means learning to translate back and forth between the various possible representations (word description, equation(s), graphs, pictures, {\dots}) of a physical situation.
\bibitem{Hispanic} We include both Mexican/Mexican American as well as Latino/Latina American under the more general category Hispanic American.  We find that Hispanic Americans made up 84\% of underrepresented students in the cohort who took Physics 9A this particular quarter and African Americans made up 11\% of this group.
\bibitem{Likert} As usual, the answers on the 5-point Likert scale (strongly agree, agree, neutral, disagree, and strongly disagree) were condensed to agree, neutral, and disagree answers.
\bibitem{OtherCLASS} One could use the individual scores, favorable and unfavorable, or even use the various subcomponents of the CLASS as independent covariates.  The decision to use a single variable was done to keep the analysis as simple as possible.  We haven't tried using all possible CLASS variables as covariates but, generally, have found that using more of these independent covariates can decrease error estimates but does not substantially change this paper's conclusions and so
is unnecessary.
\bibitem{WCGains} FCI gains were initially calculated in this student-level way so that we could analyze these gains.  If we used a class-level normalized gain\cite{Hake}, then the whole class gains would be 0.38 for AL\&CF, 0.34 for ALOnly, and
0.25 for NoAL and the UndRpMns gains would be 0.41 for AL\&CF, 0.32 for ALOnly, and 0.21 for NoAL.
\bibitem{FExamEquit} Not surprisingly, final exam grades for underrepresented minority groups are similarly statistically indistinguishable from the rest of the class if and only if the course is AL\&CF.
\end{thebibliography}
\end{document}